\documentclass[aps,reprint,twocolumn,superscriptaddress,amsmath,amssymb,floatfix,apsrev,longbibliography]{revtex4-1}
\usepackage{tabularx}
\usepackage{ gensymb,wasysym}
\usepackage{bm}
\usepackage{epsfig,subfigure}
\usepackage{multirow}
\usepackage{graphicx}
\usepackage{color}
\usepackage{amsfonts}
\usepackage{exscale}
\usepackage{amsbsy}
\graphicspath{{Fig/},{../Fig/}}
\usepackage[colorlinks,urlcolor=blue,linkcolor=blue,anchorcolor=blue,citecolor=blue]{hyperref}
\newcommand{\tabincell}[2]{\begin{tabular}{@{}#1@{}}#2\end{tabular}}

\begin{document}
\title{Realizing and  Adiabatically Preparing Bosonic Integer and Fractional  Quantum Hall states in Optical Lattices}
\author{Yin-Chen He}
\email{Corresponding author: yinchenhe@g.harvard.edu}
\affiliation{Department of Physics, Harvard University, Cambridge MA 02138}
\author{Fabian Grusdt}
\affiliation{Department of Physics, Harvard University, Cambridge MA 02138}

\author{Adam Kaufman}
\affiliation{Department of Physics, Harvard University, Cambridge MA 02138}

\author{Markus Greiner}
\affiliation{Department of Physics, Harvard University, Cambridge MA 02138}

\author{Ashvin Vishwanath}
\affiliation{Department of Physics, Harvard University, Cambridge MA 02138}
\date{\today}

\begin{abstract}
 We study the ground states of 2D lattice bosons in an artificial gauge field.
 Using state of the art DMRG simulations we obtain the zero temperature phase diagram for hardcore bosons at densities $n_b$ with  flux $n_\phi$ per unit cell, which determines a filling $\nu=n_b/n_\phi$. We find several robust quantum Hall phases, including (i) a bosonic integer quantum Hall phase (BIQH) at $\nu=2$, that realizes an interacting symmetry protected topological phase in 2D  (ii) bosonic fractional quantum Hall phases including robust states at $\nu=2/3$ and a Laughlin state at $\nu=1/2$. The observed states correspond to the bosonic Jain sequence ($\nu=p/(p+1)$) pointing towards an underlying composite fermion picture.  In addition to identifying Hamiltonians whose ground states realize these phases, we discuss their preparation beginning from  independent chains, and ramping up interchain couplings. Using time dependent DMRG simulations, these are shown to reliably produce  states close to the ground state for experimentally relevant system sizes. Besides the wave-function overlap, we utilize a simple physical signature of these phases, the non-monotonic behavior of a two-point correlation, a direct consequence of edge states in a  finite system, to numerically assess the effectiveness of the preparation scheme.  Our proposal  only utilizes existing experimental capabilities.
\end{abstract}

\maketitle

The two-dimensional Bose-Hubbard model is one of the simplest many body systems that exhibits nontrivial  physics.
Initially proposed as a model for the superconductor insulator transition in solid state system~\cite{Fisher1989,Sondhi1997}, it was later realized most cleanly in optical lattices of ultracold atoms~\cite{Greiner2002, Zhang2012}. 
It has been widely studied by varying the ratio of hopping to  interaction strength $t/U$, and the filling $n_b$ of bosons per site, which reveals the superfluid and Mott insulator phases and the quantum phase transition between them. 
A third natural parameter is the magnetic flux $n_\phi$, tuning of which has been demonstrated recently in ultra-cold atomic systems in periodically driven optical lattices~\cite{ Aidelsburger2013,Atala2014, Aidelsburger2015, Tai2016}. 
The phase diagram as a function of magnetic flux through the unit cell is less understood. This is the bosonic analog of the Harper-Hofstadter problem of free electrons in a tight binding model with magnetic flux \cite{Hofstadter1976}. However, the bosonic problem is necessarily interacting and consequently allows for a richer variety of phases (also see related study of fractional Chern insulator  \cite{Neupert2011, Regnault2011,Tang2011, Sheng2011,Grushin2014}). 

At finite flux density, quantum Hall phases \cite{Klitzing1980, Tsui1982, Laughlin1983} of bosons might appear if the filling factor $ n_b/n_\phi$ is appropriate.
In the continuum limit  $U\sim n_\phi\ll 1$ where the physics of LL applies, it was established numerically (e.g. see a review \cite{Cooper2008}) and analytically~\cite{Harper2014,Scaffidi2014}  that quantum Hall states appear.
Many of those quantum Hall phases can be understood simply using Jain's composite fermion approach \cite{Jain1989}.
For example, one can first attach one flux quanta to the boson, converting them into composite fermions, and letting them form a $\nu_{\textrm {CF}}=p$ integer quantum Hall state. This construction gives a quantum Hall state with filling factor $\nu=p/(p+1)$, which is called Jain sequence states.
Among those sequence states, there is a special one at $\nu=2$ (with $p=-2$), whose existence was overlooked for decades.
The $\nu=2$ state is called bosonic integer quantum Hall state (BIQH) \cite{ Lu2012, Senthil2013}, it belongs to the newly discovered symmetry protected topological (SPT) phase \cite{Chen2013, Haldane1983,Pollmann2010},  different from all other states that are intrinsically  topologically ordered. 
This BIQH state was however not found in the continuum limit, unless one turns to a more complicated setup, e.g. two-component bosons or higher  Chern number flat bands \cite{Furukawa2013,Geraedts2013,Regnault2013,Wu2013,Sterdyniak2015,He2015, Moeller2015, Zeng2016}. 

The absence of BIQH state in the continuum limit makes it more interesting to pass to the lattice,  on which one can achieve the infinite interaction limit $U/t\rightarrow \infty$ that may not be continuously connected with the continuum limit. Indeed, early work motivating the search for lattice effects reported a candidate BIQH at low densities ($n=1/7,\,1/9$) \cite{Moeller2009}. 
Also, previous ED calculations on small system size found several Jain's sequence states with small $p$ (e.g. $p=1$) \cite{Sorensen2005, Moeller2009, Palmer2008, Hafezi2007, Palmer2006,Sterdyniak2012, Sterdyniak2015a}.
 It is important to know how far one can go beyond the dilute limit and how large can $p$  be pushed. For example can one realize in a simple fashion the BIQH state (at $p=-2$), which would be one of the few realistic routes to realizing an  SPT phase of bosons in 2D?
This is the first question that we will address in the paper.

Even if a quantum Hall state is  the groundstate of a simple Harper-Hofstadter model, it remains challenging for cold-atom experiments to realize. Cooling into a nontrivial ground state poses special challenges particularly in the context of driven systems such as the Floquet engineered optical flux lattice systems \cite{Aidelsburger2013,Atala2014, Aidelsburger2015, Tai2016, Jaksch2003, Dalibard2011, Cooper2011, Gerbier2010, Kolovsky2011, Goldman2014}. 
To overcome this issue, a clever cooling scheme is demanding.
One way of cooling, called adiabatic preparation~\cite{Sorensen2010,Popp2004,Grusdt2014,Barkeshli2015}, begins with a trivial state with low entropy, which is then slowly ramped to the desired final state. 
Such adiabatic preparation schemes in general require a continuous phase transition between the initial state and the final state.
For a quantum Hall state, an adiabatic preparation scheme is even more difficult, since usually an exotic topological phase transition will be involved~\cite{Barkeshli2015}.
Finding an appropriate adiabatic preparation scheme for optical lattice quantum Hall states  is the second question on which we will make progress, and in particular our scheme appears to work for most quantum Hall phases, at least for the system sizes relevant for experiments.

We will first present our DMRG simulation \cite{White1992,White1993,McCulloch2008} which numerically finds robust Jain sequence states $p/(p+1)$ (e.g. $p=1, \pm 2, \cdots, \pm 5$) on the lattice with a relatively high particle density.
In particular, the BIQH state (at $p=-2$) is found robust with a short correlation length and quantized Hall conductance. A related state was observed in the low density limit in Ref. \cite{Moeller2015}.
Next we use time-dependent DMRG simulations~\cite{Vidal2003, White2004, Daley2004, Zaletel2015_time_evolution} 
as well as exact diagonalization
to discuss the adiabatic preparation scheme for quantum Hall phases, focusing on the BIQH state.
The basic idea is beginning with the independent chain limit of 1D Luttinger liquids, and ramping up interchain couplings that also introduce the flux. 
To benchmark the effectiveness of our preparation scheme, we utilize the wave-function overlap between the state generated by the time ramp and the true ground state as an indicator. 
We also discuss a physical diagnosis using two-point correlation function to detect the gapless edge state of quantum Hall phases.

\emph{Model and Phases.---}We consider the Bose-Hubbard model (Harper-Hofstadter model) on a square (triangular) lattice,
\begin{equation}\label{eq:model}
H=- J\sum_{\langle i j \rangle} e^{i A_{ij}} a^\dag_i a_j + U \sum_i  n_i(n_i-1)
\end{equation}
The first term is the nearest neighbor hopping subject to a background flux $A_{ij}$, with $\sum A=\phi=2\pi n_\phi$ on each square plaquette (or $n_\phi/2$ on each triangle). 
The second term is the on-site Hubbard interactions, and we mainly consider the limit $U\rightarrow \infty$ that gives the hard-core boson constraint $n=0, 1$.

\begin{figure}
\includegraphics[width=0.45\textwidth]{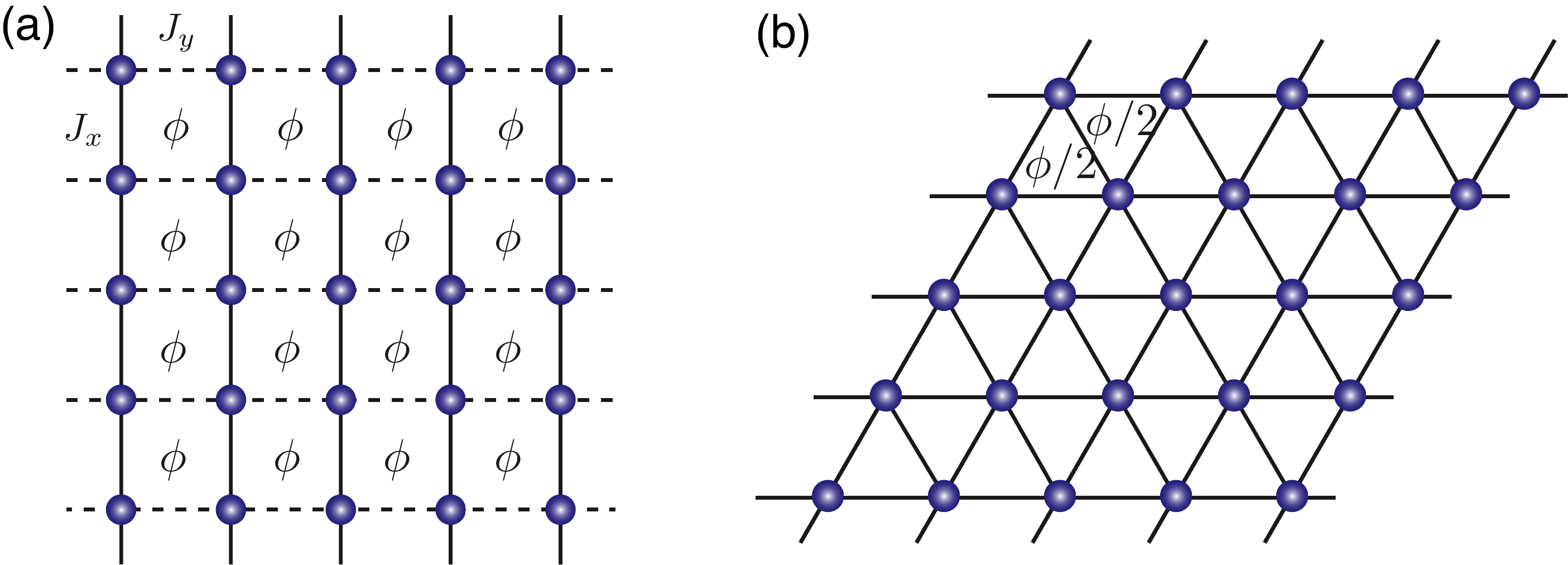}\caption{Harper-Hofstadter model on  (a) square lattice with flux $\phi=2\pi n_\phi$ on each square plaquette, (b) triangular lattice with flux $n_\phi/2$ on each triangle plaquette.}
\end{figure}

One may expect quantum Hall phases for certain filling factor $\nu=n_b/n_\phi$, where $n_b$ is the boson density per site.
The simplest possibility is the Jain sequence with $\nu=p/(p+1)=n_b/n_\phi$ from the  composite fermion approach~\cite{Jain1989}.
First, one can attach one flux quanta to the boson, yielding the composite fermion. 
The composite fermions still have density $n_b$ and  see an effective flux $n_\phi- n_b=n_b/p$, 
then naturally they will form an integer quantum Hall state with $\nu_{\textrm{CF}}=p$.
Naively, the continuum limit, which can be formulated as  lowest Landau level with contact Haldane's pseudo-potential $V\delta(r -r')$, is the most ideal platform for quantum Hall phases.
In that limit, however, several states particularly the BIQH state ($p=-2$) were not found in the extensive study (e.g. see a review \cite{Cooper2008}).

In this paper, we focus mainly on the limit with $U\rightarrow \infty$. Unexpectedly we numerically find that the Jain sequence states are more stable than in the continuum limit, in particular the BIQH state can appear in a simple setup without going to two-components  bosons or higher Chern number flat bands that previous studies focused on \cite{Furukawa2013,Geraedts2013,He2015,Regnault2013,Sterdyniak2015,Wu2013, Moeller2015, Zeng2016}.
We also note that even if $n_\phi\ll 1$,  the system we consider here is still different from the continuum limit.
It is because the infinite on-site interaction $U$ will be much larger than the Landau level spacing ($\sim n_\phi J$), which may cause strong mixing of Landau levels.

\begin{table}
\caption{\label{Table:summary} A brief summary of Jain's sequence on the square lattice with small $p=1, \pm 2$ obtained in our DMRG simulations.
$n_b$ is the density per site. $n_\phi$ is the flux per  square plaquette.
The simulations are mainly carried on an infinite cylinder with circumference  $L=6, \cdots, 12$.  
}

\setlength{\tabcolsep}{0.45cm}
\renewcommand{\arraystretch}{1.4}
\begin{tabular}{cccc}
\hline\hline
$\sigma^{xy}=\frac{p}{p+1}$ & $n_\phi$ & $n_b$  \\ \hline
\multirow{4}{*}
{\tabincell{c}{$p=1$ \\ $\sigma^{xy}=1/2$\\Laughlin State} } 			
						&$1/4$ & $1/8$ \\ 
					 	& $1/5$ & $1/10$ \\ 
						& $1/6$ &  1/12  \\ 
						& $\cdots$ & $\cdots$ \\
						\hline 
\multirow{4}{*}
{\tabincell{c}{$p=2$\\ $\sigma^{xy}=2/3$\\ Halperin's $(221)$ State}} 							
						&$1/4$ & $1/6$   \\
						&$1/5$ & $2/15$   \\
						&$1/6$ & $1/9$\\
						& $\cdots$ & $\cdots$ \\
						\hline
\multirow{4}{*}
{\tabincell{c}{$p=-2$ \\ $\sigma^{xy}=2$\\Bosonic Integer \\Quantum Hall}} 									
						& $1/6$ & $1/3$ \\ 
					 	& $1/8$ & $1/4$ \\ 
					 	& $1/10$ & $1/5$ \\ 
						& $\cdots$ & $\cdots$ \\												
						\hline\hline
\end{tabular}
\end{table}

Several methods were applied to study this problem before~\cite{Sorensen2005, Moeller2009, Palmer2008, Hafezi2007, Palmer2006,Sterdyniak2012, Natu2016,Huegel2016}, here we will  use the infinite DMRG simulation  \cite{McCulloch2008} to tackle it.
We numerically observe Jain sequence states of bosons at filling factor $\nu=p/(p+1)$ for $p=1, \pm 2, \cdots, \pm 5$.
Generally the instability of the Jain's states grows with $p$.
A consequence is that, to realize a larger $p$ one needs a more dilute density (meaning a smaller $n_\phi$ and $n_b$). 
On the other hand, we also find that the Jain sequence states are more stable on the triangular lattice (see the Table \ref{Table:triangle} in the appendix).
Here and the following we mainly focus on small $p=1, \pm 2$ on the square lattice as summarized in  Table \ref{Table:summary}. 
The results of larger $p$ are summarized in the supplementary materials.
We study infinite cylinder with circumference $L_c=6, \cdots 12$ and different sizes give consistent results.
For a smaller flux density $n_\phi$ than we show in the Tables, we expect the same quantum Hall state still exists.
A particularly interesting state corresponds to $p=-2$, that is the BIQH state at $\nu=2$. 
Unlike fractional QH, BIQH doesn't possess topological order, instead it is a SPT (protected by the $U(1)$ charge conservation).
So here our results provide a very simple setting for experimentally realizing the putative interacting SPT phase in spatial dimension higher than $d=1$.
We  note that Ref.~\cite{Huegel2016} suggested an SPT phase at $n_\phi=1/4, n_b=1/2$ with an anisotropic hopping, but we find its Hall conductance is always $0$.

\begin{figure}
\includegraphics[width=0.45\textwidth]{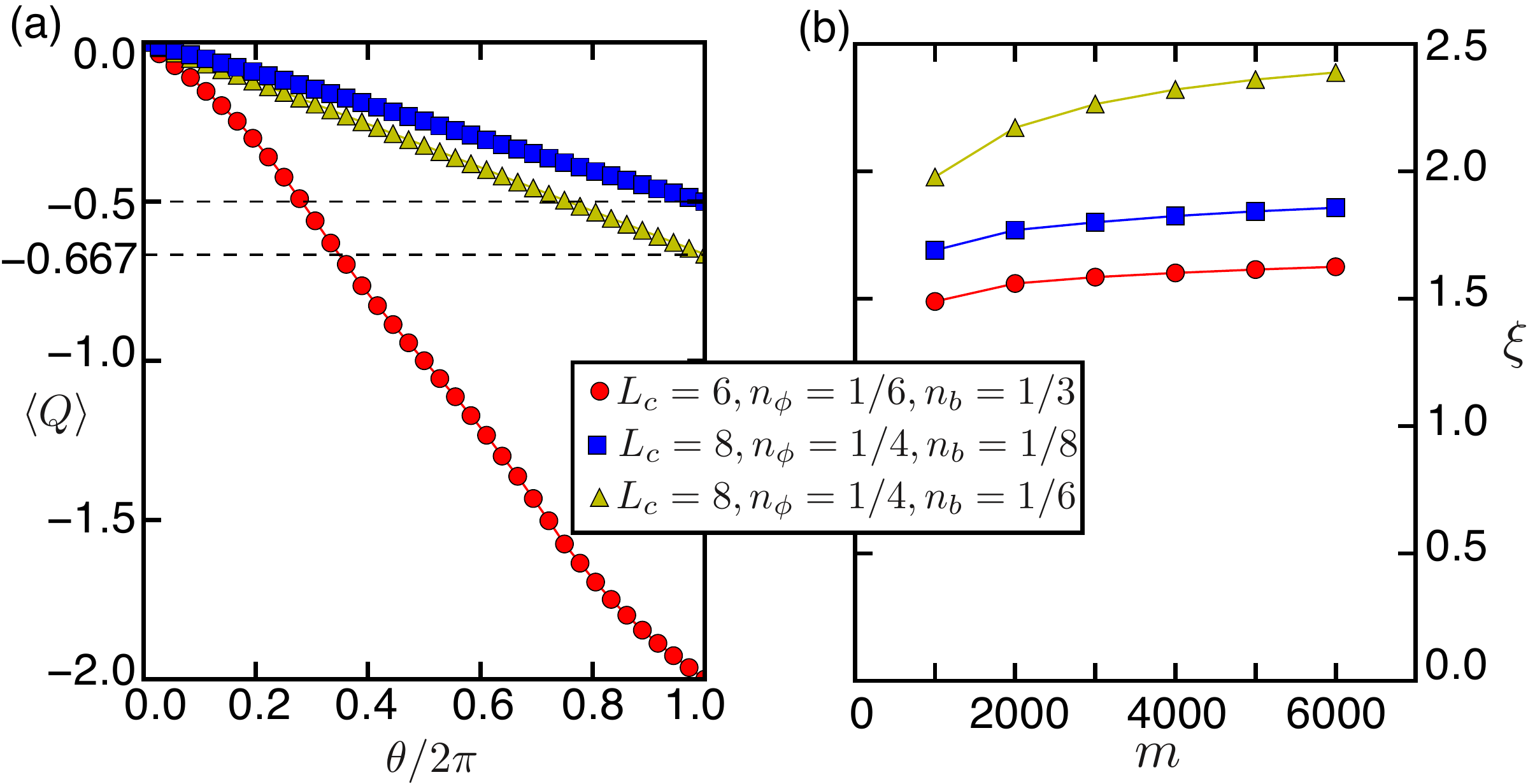}\caption{\label{fig:QH_diagnosis}Numerical diagnosis of quantum Hall states $\nu=2, 1/2, 2/3$, $L_c$ is the circumference. (a) Quantized Hall conductance measured from flux insertion on an infinite cylinder. Charge transfer as a function of flux: $\langle Q\rangle =-\sigma_{xy}\frac\theta{2\pi}$. (b) The correlation length $\xi$ of quantum Hall state versus bond dimension $m$ in DMRG simulations showing convergence. The truncation error of DMRG simulation is around $10^{-8}\sim 10^{-10}$.
 }
\end{figure}

We numerically diagnose  those quantum Hall phases through their quantized Hall conductance (many-body Chern number).
To measure the Hall conductance, we wrap the system on a cylinder, and measure the charge pumping by threading $2\pi$ flux~\cite{He2014a, Grushin2015}.
The pumped charge is exactly the Hall conductance $\sigma^{xy}$~\cite{Laughlin1981}.
As clearly shown in Fig. \ref{fig:QH_diagnosis} (a), the Hall conductance is precisely $\sigma^{xy}=1/2, 2/3, 2$ for three quantum Hall states.
Also we find that the states  have a short correlation length (Fig. \ref{fig:QH_diagnosis} (b)), indicating a fully gapped state.

\emph{Adiabatic preparation from the 1D phase.---}One important challenge for cold atom experiments is to cool into the ground state. We now discuss one preparation scheme for preparing quantum Hall phases using  adiabatic preparation starting from decoupled 1D wires.
The idea is that we first turn off hopping along one direction (say $J_y=0$).
In this limit, we get decoupled 1D Luttinger liquids  with density $n_b$.
Then we slowly turn on the hopping $J_y$ (that also introduces the flux), which eventually yield a 2D bosonic quantum Hall phase at the isotropic limit $J_y=J_x$.

Numerically we find this scheme can achieve the adiabatic preparation for bosonic  (both fractional and integer) quantum Hall phases. 
One piece of numerical evidence is the properties of the groundstate as we ramp the system from 1D wires to a 2D quantum Hall states.
First we find the physical quantities (e.g. the energy and entanglement entropy) evolves continuously as we change the parameter ($J_y$).
Second we observe that the wave-function of the groundstate of the system is changing smoothly, namely  the wave-function overlap  $|\langle \psi(J_y)|\psi(J_y+dJ_y)\rangle|\rightarrow 1$ as $d J_y\rightarrow 0$.

\begin{figure}
\includegraphics[width=0.45\textwidth]{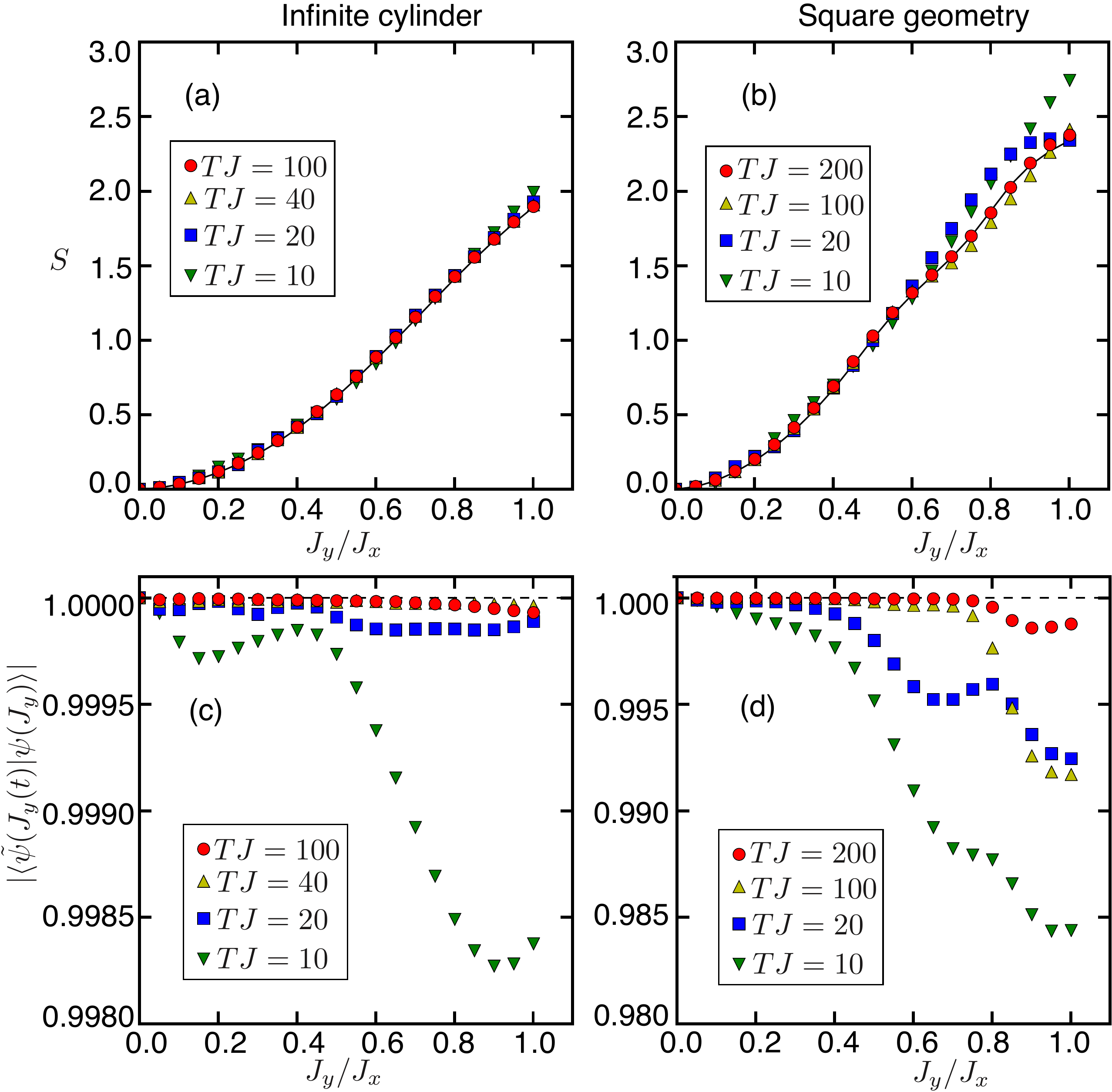}\caption{\label{fig:schemeB} Non-equilibrium dynamics simulation of the preparation scheme  with different ramp time $T$. We show the results of the BIQH state at $n_\phi=1/6$, $n_b=1/3$ of a square lattice placed on both the $L_c=6$ infinite cylinder (a), (c) and the $6\times 6$ square geometry (b), (d).
(a), (b) shows  the time evolution of entanglement entropy, where the solid line represents the entanglement entropy of the groundstate versus $J_y/J_x$, the dots represent the entanglement entropy from the time evolution. (c), (d) shows the wave-function overlap per-site between the groundstate $\psi(J_y)$ and the state from time evolution $\tilde \psi(J_y(t))$.}
\end{figure}

To make a more direct contact with experiments, we also simulate the preparation scheme as the non-equilibrium process.
It can be generally described by 
%\begin{equation}
$|\psi_f\rangle=\int_{0}^{T} dt e^{-itH(t)}|\psi_0\rangle$.
%\end{equation}
$H(t)$ is the time-dependent Hamiltonian that will be tuned experimentally,
\begin{align}
H(t)=& -J_x\sum e^{i A_{ij}} a^\dag_i a_j -J_y(t)\sum e^{i A_{ij}} a^\dag_i a_j \nonumber \\
&+ U \sum_i  n_i(n_i-1)
\end{align}
with time-dependent hopping on the $y$ direction, $J_y(t)=J_x t/T$. 
$\psi_0$ is the initial state, that is the groundstate of the starting Hamiltonian $H(0)$.
Numerically we first discretize the time-evolution operator,
$\int_{0}^{T} dt e^{-it H(t)}\approx \prod_{n=0}^m  e^{-i(t_{n+1}-t_n)H(t_{n+1})}$,
with $t_{n}=nT/m$, $m\gg 1$, and the final Hamiltonian $H(t_f=T)$ is the one in Eq. \eqref{eq:model}.
Following the method introduced by  Zaletel, et al. \cite{Zaletel2015_time_evolution}, we then rewrite the operator $e^{-i(t_{n+1}-t_n)H(t_{n+1})}$ as a matrix product operator, and apply it to the wave-function successively.

Fig. \ref{fig:schemeB} shows the numerical results for the preparation scheme.
We carry on simulations for i) the infinite cylinder geometry ($y$ direction is taken to infinite); ii) the finite square geometry that has open boundary condition on both the $x$ and $y$ direction.
To quantify how good the adiabatic preparation is, we compare the state from the time-evolution ($\tilde \psi(J_y(t))$) with the true groundstate ($\psi(J_y)$) of the static Hamiltonian.
Specifically, we compare the entanglement entropy and wave-function overlap between two states.
Clearly the adiabatic preparation works well for both schemes, and the quality of the adiabaticity increases as the preparation time becomes longer.
In particular the wave-function overlap (per site) can reach $0.9999$, which is a strong proof for our adiabatic preparation scheme.

The finite square geometry works much worse than the infinite cylinder geometry (e.g. see Fig. \ref{fig:schemeB}(c)-(d)).
Such behavior is generically expected since the quantum Hall state on a finite square geometry has gapless edge modes.
The existence of gapless modes will inevitably lead to some undesired excitations in an adiabatic preparation scheme.
Fortunately, the experimental study as well as our numerical simulations are carried out on a finite system, which has a finite gap $\Delta E\propto 1/L$.
Therefore as long as the ramping time is long enough, the adiabatic preparation can be ideally achieved.
The preparation scheme can be further optimized by adding a second tuning parameter, the magnetic flux $\phi=2\pi n_\phi$ \cite{Grusdt2014}.
Tuning of $\phi$ has recently been realized using quantum gas microscopes \cite{Tai2016}.  
More details can be found in the  Appendix (Sec. \ref{sec:opt}).

\begin{figure}
\includegraphics[width=0.45\textwidth]{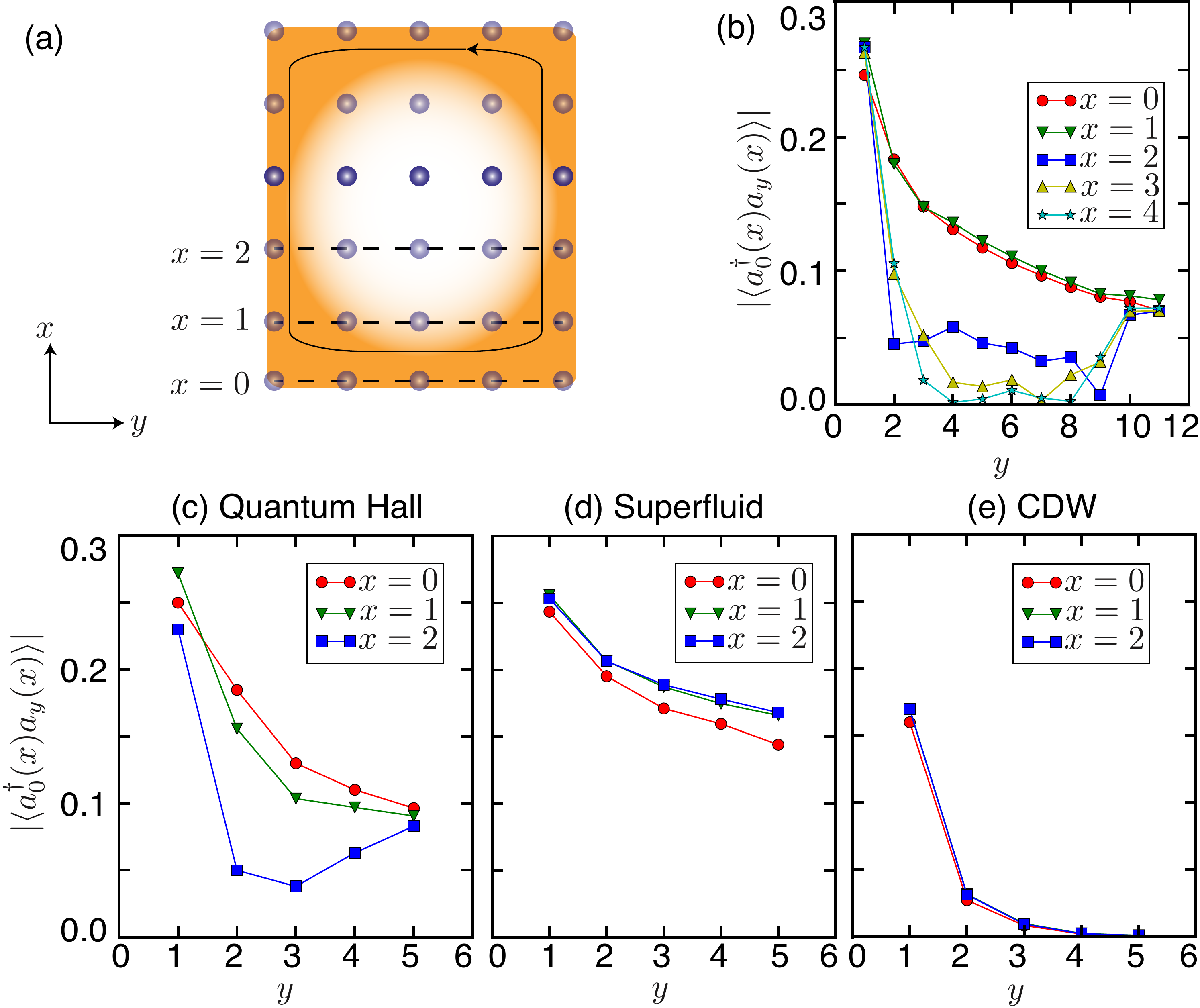} \caption{\label{fig:measurement}Diagnosis of quantum Hall state by  measuring two point correlation function $|\langle a^\dag_0(x) a_y(x) \rangle|$.
(a) The cartoon picture for a quantum Hall state on a finite  $L_x\times L_y$  cluster. 
For a quantum Hall state, $x=0, 1$ corresponds to the edge on which the correlation function decays algebraically. $x>1$ corresponds to the bulk where the correlation function decays exponentially, however if $a_r(x)$ hits the edge ($r\sim L_y$), the correlation function obeys power law.
Numerical results:
(b) $9\times 12$ cluster, $n_\phi=1/6$, $n_b=1/3$, $\nu=2$ quantum Hall state.
(c) $6\times 6$, $n_\phi=1/6$, $n_b=1/3$, $\nu=2$ quantum Hall state.
(d) $6\times 6$, $n_\phi=0$, $n_b=1/3$, superfluid.
(e) $6\times 6$, $n_\phi=1/6$, $n_b=1/2$, staggered potential $\Delta=2$, charge density wave.}
\end{figure}

\emph{Physical diagnosis   of quantum Hall state.---} Finally we study a simple correlation function based method to diagnose quantum Hall states in mesoscopic geometries. Although it is presently unclear how to directly measure this quantity in experiments, this will serve as a proxy for other correlation function based approaches to study quantum Hall states. 
The QH state has a gapped bulk but a gapless edge.
To observe this property, one can measure the correlation function $\langle a_0^\dag(x) a_y(x)\rangle$ along one direction as shown in Fig. \ref{fig:measurement}(a). 
$x$ represents the position on the $\vec x$ direction, and $a_0(x)$ is always placed on the edge.
When $x\sim 0$, the two-point correlation function is always measured on the edge, hence will give a power law decaying behavior $\langle a^\dag_0 (x) a_y (x)\rangle \propto 1/y^\alpha$.
On the other hand, when $x$ is placed in the middle of the sample ($x\sim L_x/2$), $\langle a^\dag_0 (x) a_y (x)\rangle $ is measuring the correlation function in the bulk yielding an exponentially decay behavior $e^{-y/\xi}$.
However, once $a_r(x)$ hits the edge ($r\sim L_y$), $\langle a^\dag_0 (x) a_y (x)\rangle $ will follow a power law decay again.
In summary, the two-point correlation functions behave as,
\begin{equation} \label{eq:scaling}
\langle a^\dag_0 (x) a_y (x)\rangle \propto \left\{  
\begin{aligned}
&1/y^\alpha, \quad\quad\,\, x\sim 0 \\
&e^{-y/\xi}, \quad\quad x\sim L_x/2, y<L_y\\ 
&1/(y+2x)^\alpha, \,\, x\sim L_x/2, y\sim L_y \\
\end{aligned}
\right.
\end{equation}

Fig. \ref{fig:measurement}(b) shows data of two point correlation functions of the bosonic integer quantum Hall state on a large system, $9\times12$ cluster.
It is consistent with the above scaling behavior, Eq. \eqref{eq:scaling}.
In particular, when $x\sim L_x/2$, the correlation function shows a non-monotonic behavior, it at first decays fast, but then suddenly increases as $a_y$ hits the edge.
Such scaling behavior is also visible on a small system size, e.g. $6\times 6$ cluster in Fig. \ref{fig:measurement}(c).
In contrast, a superfluid (Fig. \ref{fig:measurement}(d)) and a charge-density-wave (Fig. \ref{fig:measurement}(e)) does not show any non-monotonic behavior.
The state from our adiabatic preparation protocol also admits such non-monotonic correlations (see Fig. \ref{fig:cor_preparation} in the appendix) demonstrating that it retains physical characteristics of the ground state.

\emph{Conclusion and outlook.---} We study the Harper-Hofstadter model with hardcore bosons, and numerically find a number of quantum Hall phases, in particular a bosonic integer quantum Hall state that is hard to realize in the continuum limit.
 We also propose an adiabatic preparation scheme for those phases, and numerically show its effectiveness using time-dependent DMRG simulations.
At last we describe a simple physical diagnosis for quantum Hall phases utilizing the two point correlation functions to detect edge states. We demonstrate that the state obtained by the adiabatic preparation scheme produces the same signature as the true groundstate. While this signature is challenging to measure experimentally, we note that other viable detection schemes which could characterize these states have been proposed, for example by extracting the Hall conductance from a measurement of the center-of-mass motion perpendicular to an applied force \cite{Aidelsburger2015} or edge state (e.g. \cite{Yin2016}).
An outstanding problem for future theoretical work is to come up with additional measurement protocols.  Our theoretical study on one hand lends support to Jain's composite fermions picture in a simple setting, and on the other hand indicates a way forward for the experimental study of quantum Hall phases in optical lattices.

\emph{Acknowledgement.---}We acknowledge stimulating discussions with Eugene Demler, Frank Pollmann, Chong Wang, Norman Yao, and Liujun  Zou.
 YCH thanks  F. Pollmann for sharing the Tenpy package to do the simulations of time-dependent DMRG, and thanks  N. Yao for sharing unpublished ED results.
YCH and FG are supported by a postdoctoral fellowship from the Gordon and Betty Moore Foundation, under the EPiQS initiative, GBMF4306, at Harvard University. AV is supported by a Simons Investigator Award and by the AFOSR MURI grant FA9550- 14-1-0035.

\bibliography{HH.bib}

\newpage

\appendix
\section{More numerical results}

\subsection{Jain's sequence}

We find on the triangular lattice, Jain's sequence becomes more stable. 
Table \ref{Table:triangle} lists our DMRG results for a small $p=1, \pm 2$.
Clearly for the same $p$ on the square lattice (Table \ref{Table:summary}),  the density $n_\phi, n_b$ can be larger. 

Numerically we also find  Jain's sequence states with larger $p$ as summarized in Table \ref{Table:summaryII}. 
Compared with small $p$ that we presented in the main text, the larger $p$ state is less stable.
It, on the other hand, requires larger system size for the numerical simulations.
Therefore  it is  more difficult to firmly establish the existence of those quantum Hall phases.
However, in our simulations for several system size $L_c=6, \cdots 12$, we consistently find those quantum Hall states as the groundstates.
Therefore, it is reasonable to expect those quantum Hall states are stable in the thermodynamic limit, similar as a smaller $p$.

\begin{table}\caption{\label{Table:triangle} A brief summary of Jain's sequence on the triangular lattice.
$n_b$ is the density per site. $n_\phi/2$ is the flux per  triangle plaquette.
The simulations are mainly carried on an infinite cylinder with circumference  $L=6, \cdots, 12$. 
}

\setlength{\tabcolsep}{0.45cm}
\renewcommand{\arraystretch}{1.4}
\begin{tabular}{cccc}
\hline\hline
$\sigma^{xy}=\frac{p}{p+1}$ & $n_\phi$ & $n_b$  \\ \hline
\multirow{4}{*}
{\tabincell{c}{$p=1$ \\ $\sigma^{xy}=1/2$\\Laughlin State} } 			
						&$2/5$ & $1/5$\\
						&$1/3$ & $1/6$\\
						&$1/4$ & $1/8$ \\ 
						& $\cdots$ & $\cdots$ \\
						\hline 
\multirow{4}{*}
{\tabincell{c}{$p=2$\\ $\sigma^{xy}=2/3$\\ Halperin's $(221)$ State}} 							
						&$3/8$ & $1/4$ \\
						&$3/10$ &$1/5$\\
						&$1/4$ & $1/6$   \\
						& $\cdots$ & $\cdots$ \\
						\hline
\multirow{4}{*}
{\tabincell{c}{$p=-2$ \\ $\sigma^{xy}=2$\\Bosonic Integer \\Quantum Hall}} 									
						& $1/6$ & $1/3$ \\ 
					 	& $1/8$ & $1/4$ \\ 
					 	& $1/10$ & $1/5$ \\ 
						& $\cdots$ & $\cdots$ \\												
						\hline\hline
\end{tabular}
\end{table}

\begin{table}
\caption{\label{Table:summaryII} A brief summary of Jain's sequence on the square lattice with larger $p$ ($= \pm 3, \pm 4, \pm 5$) obtained in our DMRG simulations.
$n_b$ is the density per site. $n_\phi$ is the flux per  square plaquette.
The simulations are mainly carried on an infinite cylinder with circumference $L_c=6, \cdots, 12$. }
\setlength{\tabcolsep}{0.45cm}
\renewcommand{\arraystretch}{1.4}

\begin{tabular}{cccc}
\hline\hline
$\sigma^{xy}=\frac{p}{p+1}$ & $n_\phi$ & $n_b$  \\ \hline

\multirow{2}{*}
{\tabincell{c}{$p=3$ \\ $\sigma^{xy}=3/4$}} 									
					 	& $1/5$ & $3/20$ \\ 
						& $1/6$ &  $1/8$ \\ 
						\hline 
\multirow{2}{*}
{\tabincell{c}{$p=-3$ \\ $\sigma^{xy}=3/2$}} 									
						& $1/5$ & $3/10$ \\ 
						& $1/6$ &  $1/4$ \\ 
						\hline 	
\multirow{2}{*}
{\tabincell{c}{$p=4$ \\ $\sigma^{xy}=4/5$}} 									
					 	& $1/5$ & $4/25$ \\ 
						& $1/6$ &  $2/15$ \\ 
						\hline 
\multirow{2}{*}
{\tabincell{c}{$p=-4$ \\ $\sigma^{xy}=4/3$}} 									
						& $1/6$ & $2/9$ \\ 
					 	& $1/8$ &  $1/6$ \\ 
						\hline 			
\multirow{2}{*}
{\tabincell{c}{$p=5$ \\ $\sigma^{xy}=5/6$}} 									
						& $1/6$ & $5/36$ \\ 
					 	& $1/8$ & $5/48$ \\ 
						\hline 
\multirow{2}{*}
{\tabincell{c}{$p=-5$ \\ $\sigma^{xy}=5/4$}} 									
						& $1/6$ & $5/24$ \\ 
					 	& $1/8$ & $5/32$ \\ 
																						
						\hline\hline
\end{tabular}
\end{table}

\subsection{Correlation functions in the preparation scheme}

\begin{figure}
\includegraphics[width=0.49\textwidth]{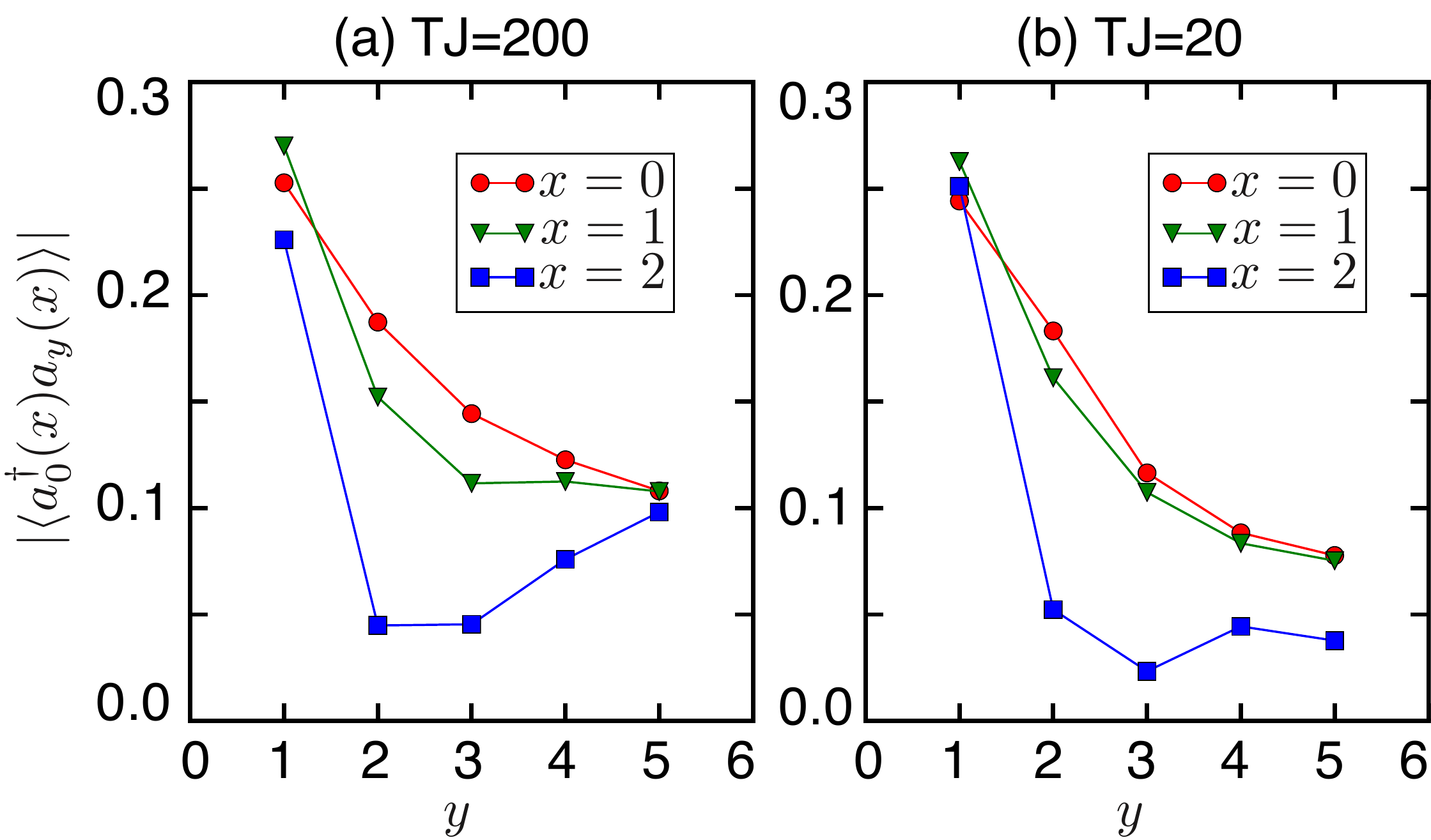} \caption{\label{fig:cor_preparation}Two-point correlation function $\langle a^\dag_0(x) a_y(x)\rangle$ in the preparation scheme. We simulate a $6\times6$ finite cluster with open boundary conditions, $n_\phi=1/6$, $n_b =1/3$. (a) $TJ=200$, (b) $TJ=20$ . }
\end{figure}

A self-consistent check of our proposal is whether the  state from our  preparation scheme also admits the scaling behavior Eq. \eqref{eq:scaling}.
Fig. \ref{fig:cor_preparation} shows the numerical data, with $n_\phi=1/6$, $n_b=1/3$ on a finite $6\times 6$ cluster.
We compare two different ramp time $TJ=200$ and $TJ=20$. Clearly the longer time ramp yields a  state with good agreement with the true ground state (Fig. \ref{fig:measurement}(c)).
On the other hand,  a shorter time ramp has a worse  performance. 

\begin{figure}
\includegraphics[width=0.49\textwidth]{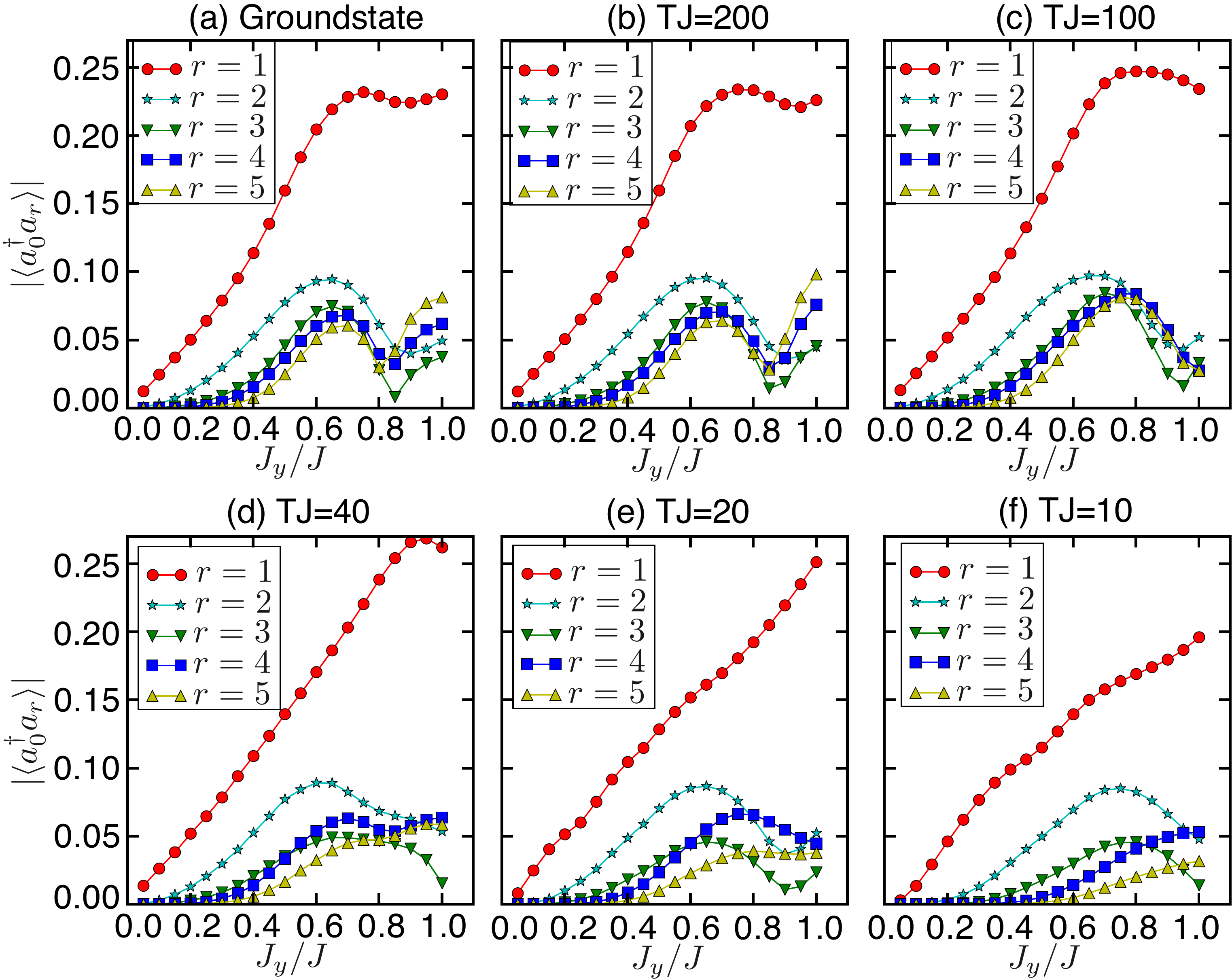} \caption{\label{fig:cor_evo_app}Two-point correlation function $\langle a^\dag_0(x=2) a_r(x=2)\rangle$ in the preparation scheme. We simulate a $6\times6$ finite cluster, $n_\phi=1/6$, $n_b =1/3$, and $x=2$. (a) Groundstate, (b) $TJ=200$, (c) $TJ=100$, (d) $TJ=40$, (e) $TJ=20$, (f) $TJ=10$. }
\end{figure}

It is also interesting to see  the behavior of correlation functions in the whole preparation process as shown in  Fig. \ref{fig:cor_evo_app}.
When $J_y=0$, the wires are decoupled hence the correlation perpendicular to the wire is vanishing. 
When $J_y$ increases, the correlation begins to develop, and it follows the monotonic behavior that it decays as $r$ increases. 
When $J_y\sim J$, the system realizes a QH state, and the correlation follows a non-monotonic behavior, namely it drops first and suddenly increases when $r$ hits the edge (as also shown in Fig. \ref{fig:measurement}).
Clearly when the ramp time is long (Fig. \ref{fig:cor_evo_app}(b)) the evolution of the correlation functions matches well with those of the groundstate (Fig. \ref{fig:cor_evo_app}(a)).
On the other hand, when the ramp time is short (e.g. Fig. \ref{fig:cor_evo_app}(f)), the correlation function has a considerable discrepancy with the groundstate when $J_y$ is large, indicating the possible failure of the adiabatic preparation.

\subsection{Optimization of preparation scheme} 
\label{sec:opt}
The preparation scheme can be further optimized by adding a second tuning parameter (besides the anisotropy $J_y/J_x$), the magnetic flux $\phi=2\pi n_\phi$ \cite{Grusdt2014}.
 Tuning of $\phi$ has recently been realized using quantum gas microscopes \cite{Tai2016}.  
In the two dimensional phase space, there are  infinite number of paths that connects the trivial phase with the quantum Hall phase.
The general principle is to find a path that has a maximum gap (finite-size gap $\Delta E \sim 1/L$).
To make a closer contact with experiments, we also consider a small system consisting of a few soft core bosons on a square lattice trapped inside a box potential. 

%%%%%%%%%%%%%%%%%%%%%%%%%%%%%%%%%%%%%%%%%%%%%%%%%%%%%
\begin{figure}
\centering
\epsfig{file=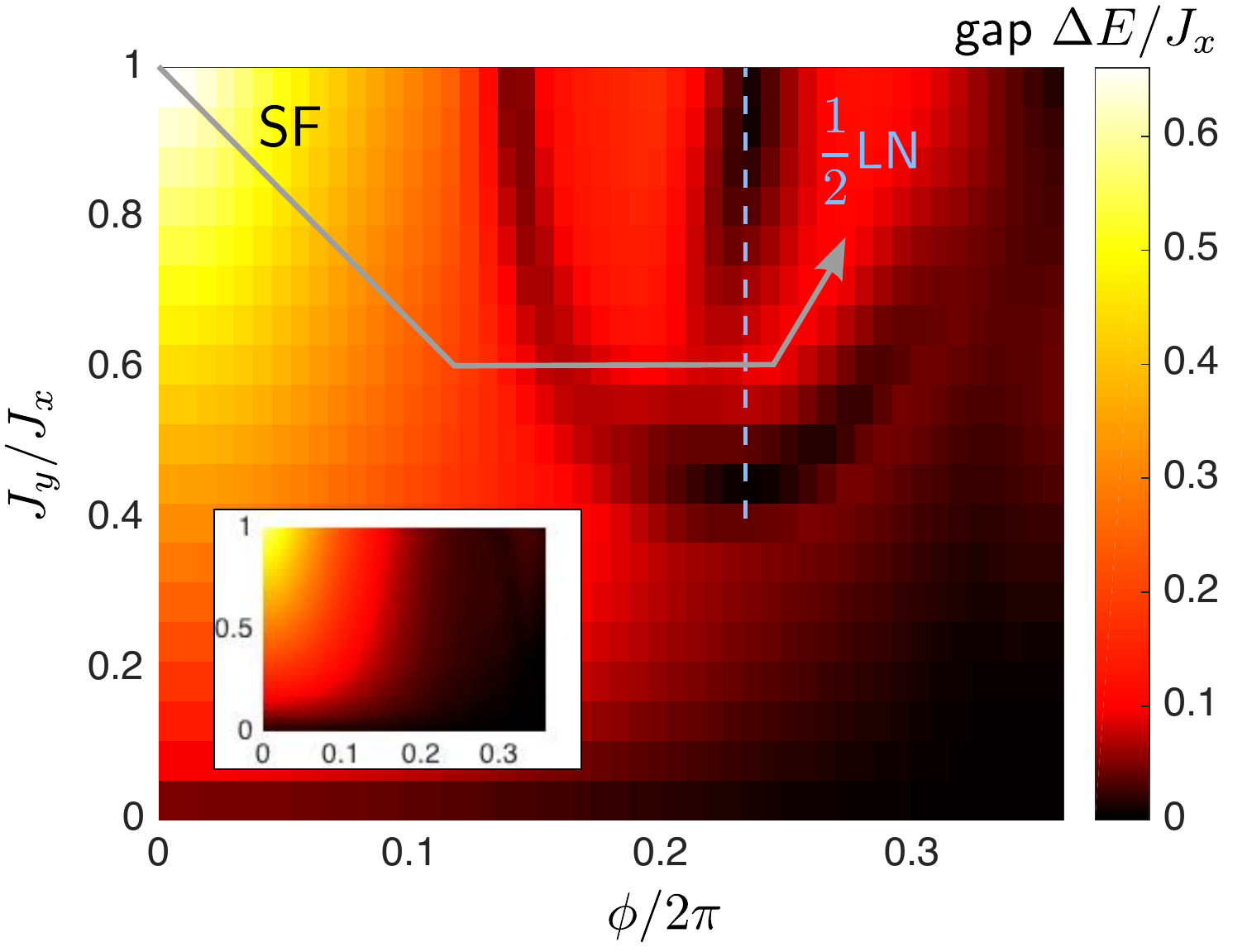, width=0.45\textwidth}
\caption{Finite-size gap $\Delta E$ for $1/2$ Laughlin states in a small box. By tuning the ration $J_y/J_x$ of the tunnel couplings, as well as the magnetic flux per plaquette, $\phi=2\pi n_\phi$, the $1/2$ Laughlin state can be adiabatically prepared starting from a superfluid (SF) phase without the magnetic field. We used exact diagonalization for $N=4$ particles on a $5 \times 6$ square lattice with open boundary conditions, with soft-core interactions of strength $U=4 J_x$. The dashed vertical line indicates where the appearance of the $1/2$ Laughlin state is expected. The inset shows the same finite-size gap for $N=1$ particle.}
\label{LaughlinAdiabaticScheme}
\end{figure}
%%%%%%%%%%%%%%%%%%%%%%%%%%%%%%%%%%%%%%%%%%%%%%%%%%%%%

Fig.~\ref{LaughlinAdiabaticScheme} shows the finite-size gap $\Delta E$ from ground to first excited state, assuming four bosons with soft-core interactions in a $5 \times 6$ square lattice. Compared to the non-interacting system (inset of the figure) we observe a finite gap when the filling $n_b/n_\phi \gtrsim N / (2N-1)$ and $J_x \approx J_y$, where a $1/2$-Laughlin state is expected \cite{Sorensen2005,Hafezi2007}. The corresponding value of the flux per plaquette is indicated by a dashed line in the figure. A promising path through parameter space with a minimum gap of $\Delta E_{\rm min}  = 0.068 J_x $ is indicated by an arrow. The finite-size gap is comparable to the bulk gap of the $1/2$ Laughlin state everywhere, which can be calculated by placing the same system on a torus with periodic boundary conditions \cite{Sorensen2005}. This establishes that bosonic Laughlin states containing a few bosons can be readily prepared in experiments with ultracold atoms \cite{Tai2016}.

\subsection{Another adiabatic preparation scheme}

\begin{figure}
\includegraphics[width=0.2\textwidth]{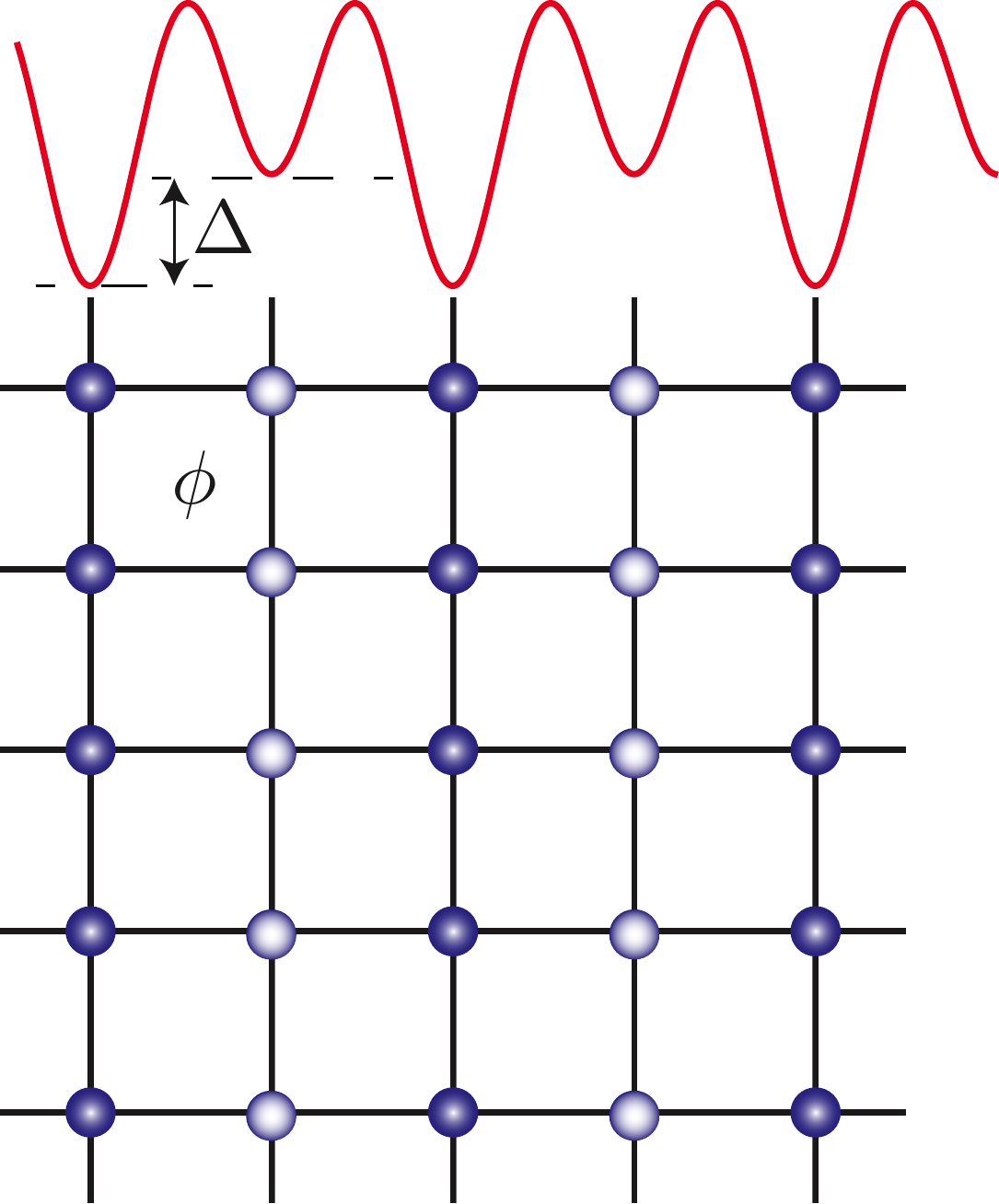}
\caption{\label{fig:scheme} The second scheme to  adiabatically prepare quantum Hall state from decoupled 1D liquid. The idea is to ramp chemical potential offset $\Delta$ between neighboring 1D wires from the infinity $\Delta\gg J$ to $\Delta=0$.}
\end{figure}

\begin{figure}[ht]
\includegraphics[width=0.49\textwidth]{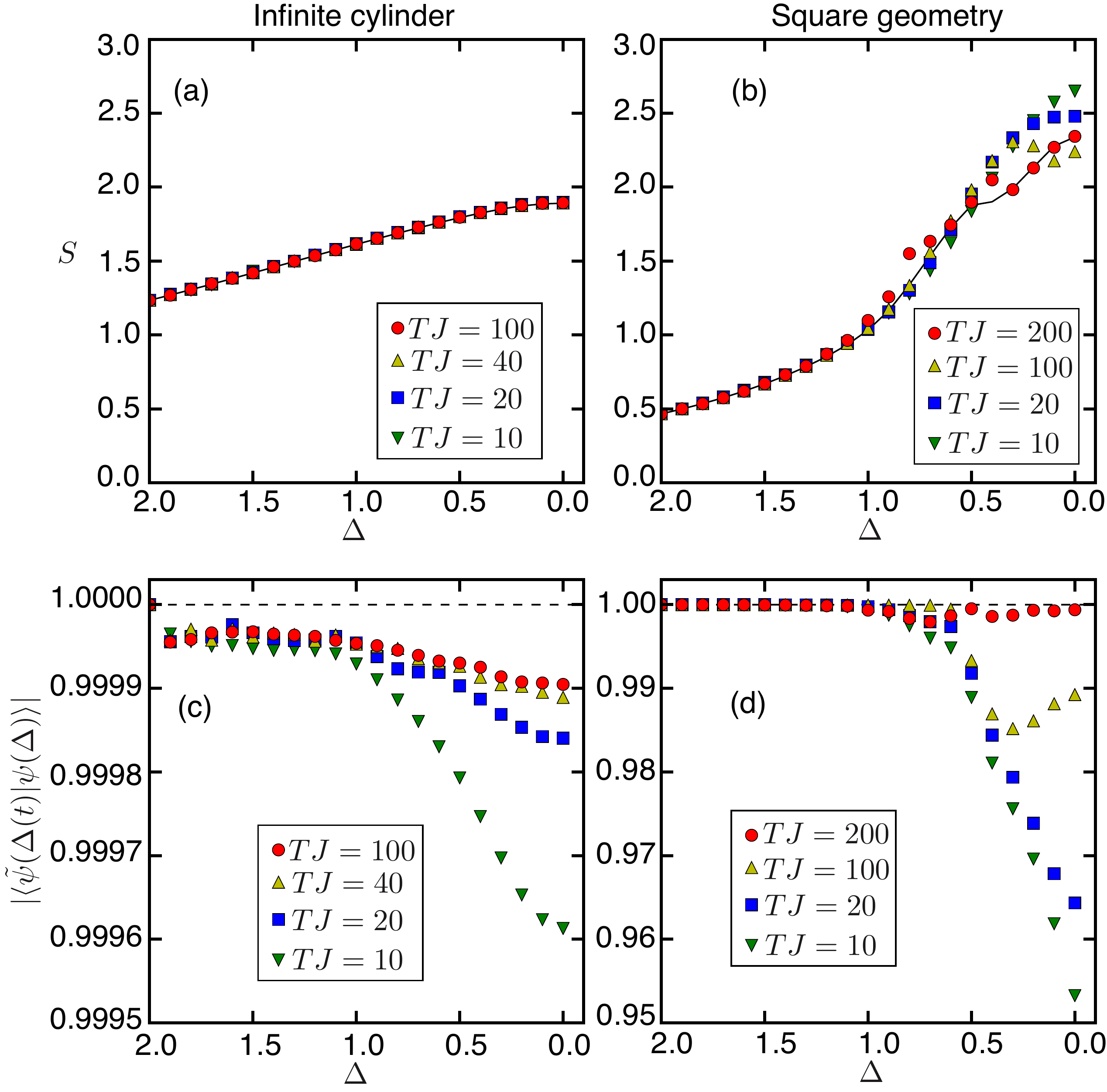}\caption{\label{fig:scheme_app} Non-equilibrium dynamics simulation of the second scheme with different ramping time $T$. We show the results of the bosonic integer quantum Hall state at $n_\phi=1/6$, $n_b=1/3$ of a square lattice placed on both the $L_c=6$ finite cylinder (a), (c) and the $6\times 6$ square geometry (b), (d).
(a), (b) shows  the time evolution of entanglement entropy, where the solid line represents the entanglement entropy of the groundstate versus $\Delta$, the dots represents the entanglement entropy from the time evolution. (c), (d) shows the wave-function overlap per-site between the groundstate $\psi(\Delta)$ and the state from time evolution $\tilde \psi(\Delta(t))$.}
\end{figure}

Besides the scheme discussed in the main text, we also find another  scheme for the adiabatic preparation.
The idea  is to begin with a system with a  staggered potential along one direction (say $y$ direction), see Fig. \ref{fig:scheme}(a). 
In particular we make the potential offset $\Delta$ between the odd and even wires to be  much larger than the hopping amplitude $J$, $\Delta\gg J$. 
Consequently, the bosons will be trapped to the odd wires, and are prohibited to hop to the neighboring wires.
Each wire is indeed described by a 1D Luttinger liquid with the density $2n_b$.
Next we slowly turn off the staggered potential,  the bosons are allowed to hop to neighboring wires.
Eventually the decoupled 1D wires will be  melt to a 2D phase, which turns out to be the bosonic quantum Hall phase.

Similar as the main text, we simulate the preparation scheme as a non-equilibrium scheme using the time-dependent Hamiltonian,
\begin{equation}
H_I(t)= -J\sum_{\langle i j \rangle} e^{i A_{ij}} a^\dag_i a_j + U \sum_i  n_i(n_i-1)-\Delta(t) \sum_{y=2k+1}  n_y
\end{equation}
where the potential on the odd wires are time-dependent
\begin{equation}
\Delta(t)=\Delta (1- t/T)
\end{equation}
where $U\gg\Delta\gg J$, $T$ is the total time for the ramp.

The numerical results are shown in Fig. \ref{fig:scheme_app}.
Similarly as the first scheme described in the main text, the preparation scheme works well as long as the ramp time is not too short.

\end{document}